\documentclass[preprintnumbers,amssymb,amsmath,superscriptaddress,letterpaper,nofootinbib,twocolumn,floatfix]{revtex4}[10pt]

\usepackage{graphicx}
\usepackage{dcolumn}
\usepackage{bm}
\usepackage{natbib}
\usepackage{calligra}
\usepackage[T1]{fontenc}
\usepackage{egothic}
\usepackage[T1]{fontenc}
\newfont{\rsfsten}{rsfs10 scaled 1200}
\newfont{\rsfsseven}{rsfs10 scaled 1200}
\newfont{\rsfsfive}{rsfs10 scaled 1200}
\usepackage{epsfig}
\usepackage{units}

\newcommand{\be}{\begin{equation}}
\newcommand{\ee}{\end{equation}}
\newcommand{\bea}{\begin{eqnarray}}
\newcommand{\eea}{\end{eqnarray}}

\def\lsim{\mathrel{\raise.3ex\hbox{$<$\kern-.75em\lower1ex\hbox{$\sim$}}}}
\def\gsim{\mathrel{\raise.3ex\hbox{$>$\kern-.75em\lower1ex\hbox{$\sim$}}}}

\begin{document}

\hspace*{130mm}{\large \tt FERMILAB-15-093-A}
\vskip 0.2in


\title{On The Gamma-Ray Emission From Reticulum II and Other Dwarf Galaxies}

\author{Dan Hooper}
\affiliation{Fermi National Accelerator Laboratory, Center for Particle Astrophysics, Batavia, IL}
\affiliation{University of Chicago, Department of Astronomy and Astrophysics Chicago, IL }

\author{Tim Linden}
\affiliation{Kavli Institute for Cosmological Physics University of Chicago, Chicago, IL }

\begin{abstract}

The recent discovery of ten new dwarf galaxy candidates by the Dark Energy Survey (DES) and the Panoramic Survey Telescope and Rapid Response System (Pan-STARRS) could increase the Fermi Gamma-Ray Space Telescope's sensitivity to annihilating dark matter particles, potentially enabling a definitive test of the dark matter interpretation of the long-standing Galactic Center gamma-ray excess. In this paper, we compare the previous analyses of Fermi data from the directions of the new dwarf candidates (including the relatively nearby Reticulum II) and perform our own analysis, with the goal of establishing the statistical significance of any gamma-ray signal from these sources. We confirm the presence of an excess from Reticulum II, with a spectral shape that is compatible with the Galactic Center signal. The significance of this emission is greater than that observed from 99.84\% of randomly chosen high-latitude blank-sky locations, corresponding to a local detection significance of 3.2$\sigma$. We improve upon the standard blank-sky calibration approach through the use of multi-wavelength catalogs, which allow us to avoid regions that are likely to contain unresolved gamma-ray sources.

\end{abstract}

\maketitle

\section{introduction}

Over the past several years, a bright and statistically significant excess of gamma-rays has been reported from the region surrounding the Galactic Center~\cite{Goodenough:2009gk,Hooper:2010mq,Hooper:2011ti,Abazajian:2012pn,Gordon:2013vta,Hooper:2013rwa,Abazajian:2014fta,Daylan:2014rsa,Calore:2014xka}. The spectral and morphological characteristics of this signal are each in good agreement with that predicted from annihilating dark matter particles with a mass of $m_{\rm DM}\sim$~35-60~GeV and a cross section of $\sigma v \sim 10^{-26}$~cm$^3$s$^{-1}$ (for the representative case of annihilations to $b\bar{b}$). And although the proposed astrophysical explanations for this excess have been shown to face considerable challenges, it is not currently possible to entirely rule out the possibility that these photons originate from a large population of unresolved point sources~\cite{Cholis:2014lta,Hooper:2013nhl,Petrovic:2014xra} or from a series of cosmic ray outbursts~\cite{Petrovic:2014uda,Carlson:2014cwa,outburst}. In light of this situation, gamma-ray observations of the Milky Way's dwarf spheroidal galaxies play a critical role, being potentially able to provide a confirmation or refutation of the dark matter interpretation of the Galactic Center excess. 

Searches for gamma-rays from known dwarf galaxies~\citep{Ackermann:2015zua,Geringer-Sameth:2014qqa,Ackermann:2013yva} have yielded stringent constraints on the dark matter parameter space. They have not yet, however, been sufficiently sensitive to cover the full range of cross sections favored to explain the Galactic Center excess. It has been anticipated that ongoing and planned optical surveys will discover a significant number of presently unknown Milky Way dwarf spheroidal galaxies~\cite{He:2013jza,Tollerud:2008ze,Hargis:2014kaa,Rossetto:2011yb}. If one of more of these objects happens to be nearby and/or contain a high density of dark matter, it could constitute an important target for the Fermi Gamma-Ray Space Telescope, significantly strengthening their sensitivity to annihilating dark matter.

Very recently, optical imaging data from the Dark Energy Survey (DES) was used to discover nine new dwarf galaxy candidates~\cite{Bechtol:2015wya,Koposov:2015cua}. Shortly thereafter, yet another dwarf candidate (Triangulum II) was discovered from within the data from the Panoramic Survey Telescope and Rapid Response System (Pan-STARRS)~\cite{PanSTARRS}. Of particular interest is the object Reticulum II (also known as DES J0335.6-5403) whose proximity ($\sim$30-32 kpc) and spatial extent (half-light radius of $55 \pm 5$ pc) make it likely to be a dwarf galaxy (rather than a globular cluster) and a very promising target for gamma-ray searches for annihilating dark matter. Although spectroscopic follow-up will be required to measure the dark matter distributions of these systems, it is plausible that Recticulum II (or perhaps Triangulum II) could provide a gamma-ray signal from annihilating dark matter that is brighter than that from any other known dwarf galaxy.  

Two independent groups have already reported the results of their analyses of Fermi Gamma-Ray Space Telescope data from the directions of Reticulium II and DES's other new dwarf galaxy candidates. 
%
The first of these analyses, presented jointly by the Fermi and DES collaborations, identified a modest gamma-ray excess from the direction of Reticulium II, with a test statistic (TS) of 6.7~\cite{Drlica-Wagner:2015xua}.\footnote{TS is defined as twice the difference in the global log-likelihood between the null and alternative hypotheses.} Although the text of that paper quotes a $p$-value for this excess of 0.06, this includes a trials factor of approximately 4, intended to account for the range of possible dark matter masses and annihilation channels that were scanned over. If one instead employs dwarf galaxies to test the dark matter interpretation of the Galactic Center excess (for which the spectrum has been previously measured), no such trials factor is required. The local significance of this signal (corresponding to $p \simeq 0.015$) is approximately 2.4$\sigma$. The other analysis, carried out by Geringer-Sameth {\it et al.}, also finds an excess from Reticulium II, and quotes significances ranging from approximately 2.3$\sigma$ to 4.7$\sigma$, depending on the background model procedure and trials factor that are adopted. The local significance (without trials for different dark matter models) determined using an empirical sample of nearby spatial regions (as opposed to assuming that the background is Poisson distributed) is found to be approximately 2.8$\sigma$~\cite{Geringer-Sameth:2015lua}, or approximately 0.4$\sigma$ greater that that found by the Fermi and DES collaborations. The excesses identified in each of these papers is most prevalent at energies between $\sim$2-10 GeV, compatible with the spectrum observed from the region surrounding the Galactic Center~\cite{Daylan:2014rsa,Goodenough:2009gk,Hooper:2011ti,Hooper:2010mq,Abazajian:2012pn,Gordon:2013vta,Hooper:2013rwa,Abazajian:2014fta,Calore:2014xka}. Considering the fact that these two groups apply quite different analysis techniques to this problem, and employ different data sets (Ref.~\cite{Drlica-Wagner:2015xua} utilizes Pass 8 data while the analysis of Ref.~\cite{Geringer-Sameth:2015lua} is restricted to the publicly available Pass 7), it is remarkable that their results are so similar. 

In this article, we present our own analysis of the (Pass 7) Fermi data from the directions of Reticulum II and other Milky Way dwarf galaxies with the goal of assessing the characteristics and statistical significance of any excess that might exist. We confirm the existence of a gamma-ray signal from Reticulum II, and assess the (local) significance of this excess to be 3.2$\sigma$. And although spectroscopic follow-up of Reticulum II will be required before this observation can be used to constrain or infer the value of the dark matter annihilation cross section, for a plausible range of dark matter profiles, this result appears to be consistent with dark matter interpretations of the Galactic Center signal and the null results from other dwarf galaxies. 

\section{Fermi Data Analysis}

\begin{figure}
\epsfig{file=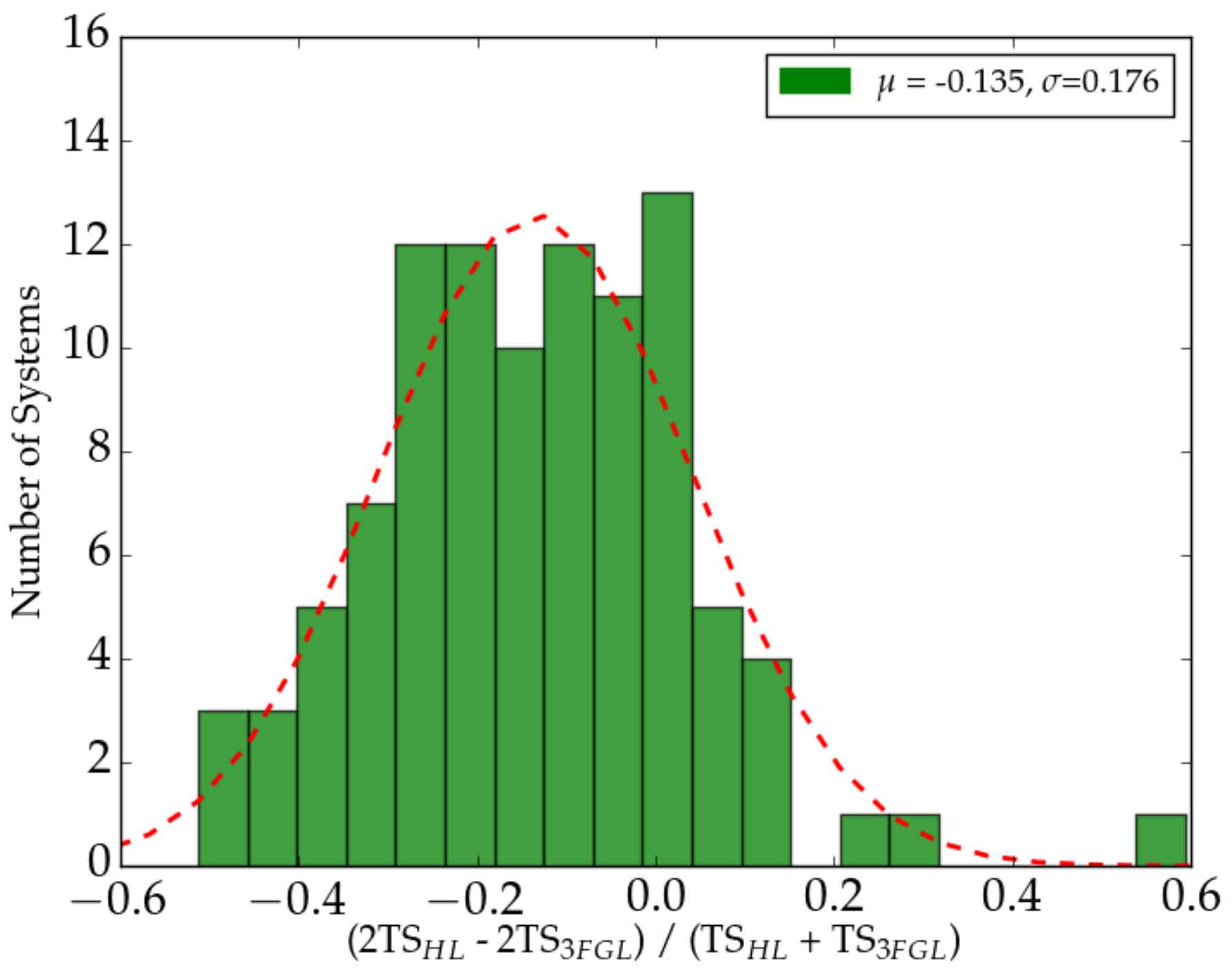,width=0.5\textwidth}   
\caption{\label{fig:3FGLTS} A comparison of the values of the test statistic found in our analysis (TS$_{\it HL}$) to those given in Fermi's 3FGL catalog (TS$_{\it 3FGL}$). Overall the agreement between these two determinations is good, validating our analysis procedure.}
\end{figure}

In order to calculate the significance of any gamma-ray emission observed from a given dwarf galaxy (or dwarf galaxy candidate), we examine approximately 6.5 years of Fermi-LAT data,\footnote{MET range: 239557417 - 447078115} using the P7REP photons in the energy range of 500~MeV to 500~GeV.  We exclude events arriving with a zenith angle greater than 100$^\circ$, as well as those which do not pass the ``Source'' photon data selection. We also exclude events that were observed while the instrument was not in science survey mode, when the instrumental rocking angle was $>$52$^\circ$, or when the instrument was passing through the South Atlantic Anomaly. For each source, we examine the photons observed within a 10$^\circ$x10$^\circ$ box centered around the location of the source, and divide the photons into 100x100 angular bins and 24 evenly spaced logarithmic energy bins. We analyze the instrumental exposure throughout the region of interest using the {\tt P7REP\_SOURCE\_V15} instrument response functions. We employ the latest model for diffuse galactic gamma-ray emission ({\tt gll\_iem\_v05\_rev1.fit}) and the latest isotropic emission template for the Source photon events ({\tt iso\_source\_v05.txt}), and include all point sources given in the 3FGL Catalog~\cite{TheFermi-LAT:2015hja}. 

In our analysis, we follow a prescription as similar as possible to that employed by the Fermi-LAT collaboration~\citep{Ackermann:2013yva, Ackermann:2015zua, Drlica-Wagner:2015xua}. Specifically, we first set the global normalization of background sources and the dwarf spheroidal over the entire 500~MeV -- 500~GeV energy range, utilizing the Fermi-LAT {\tt gtlike} code and the {\tt MINUIT} algorithm. In this phase, we seed the dwarf spheroidal spectrum as a simple power-law with an index of -2.0. We then fix the normalization of all background components, and employ the \emph{pyLikelihood} package to scan the flux of the dwarf spheroidal in each energy bin, calculating the delta-log-likelihood ($\Delta LG(L)$) as a function of the source flux. In order to calculate the TS for a given dark matter model, we minimize the total log-likelihood summed over all energy bins after constraining the photon flux by the spectral shape of the dark matter model.

To validate the results of this method, we perform two tests. First, we randomly select 100 Fermi-LAT 3FGL point sources~\cite{TheFermi-LAT:2015hja} with $|b|$~$>$~30$^\circ$, a ``curve significance'' smaller than 2 (indicating consistency with a power-law spectrum) and a TS smaller than 100 in the energy range of 300 MeV to 100 GeV. In order to compare our results to those given in the 3FGL catalog, we employ the above technique with the following modifications. We restrict our analysis to four years of Fermi-LAT data,\footnote{MET range: 239557417 - 365467563} evaluate an energy range of 300 MeV -- 100 GeV in 20 energy bins utilizing 8 energy bins per decade except for the final bin (which was extended to an energy of 100 GeV), and we scan the likelihood fits using power-law, rather than dark matter motivated, spectral shapes. In Fig.~\ref{fig:3FGLTS} we show the distribution of the TS calculated in our analysis (TS$_{\it HL}$) compared to that obtained by the Fermi-LAT collaboration (TS$_{\it 3FGL}$) in the same energy range. We find that our TS values are, on average, slightly (13.5\%) lower those reported in the 3FGL. We attribute this primarily to the fact that we normalize the background by fitting over a 10$^{\circ}\times 10^{\circ}$ region, rather than over the entire sky. The dashed curve in Fig.~\ref{fig:3FGLTS} represents the best-fit gaussian of this distribution, with a mean of -0.135 and a standard deviation of 0.176.

\begin{figure}
\epsfig{file=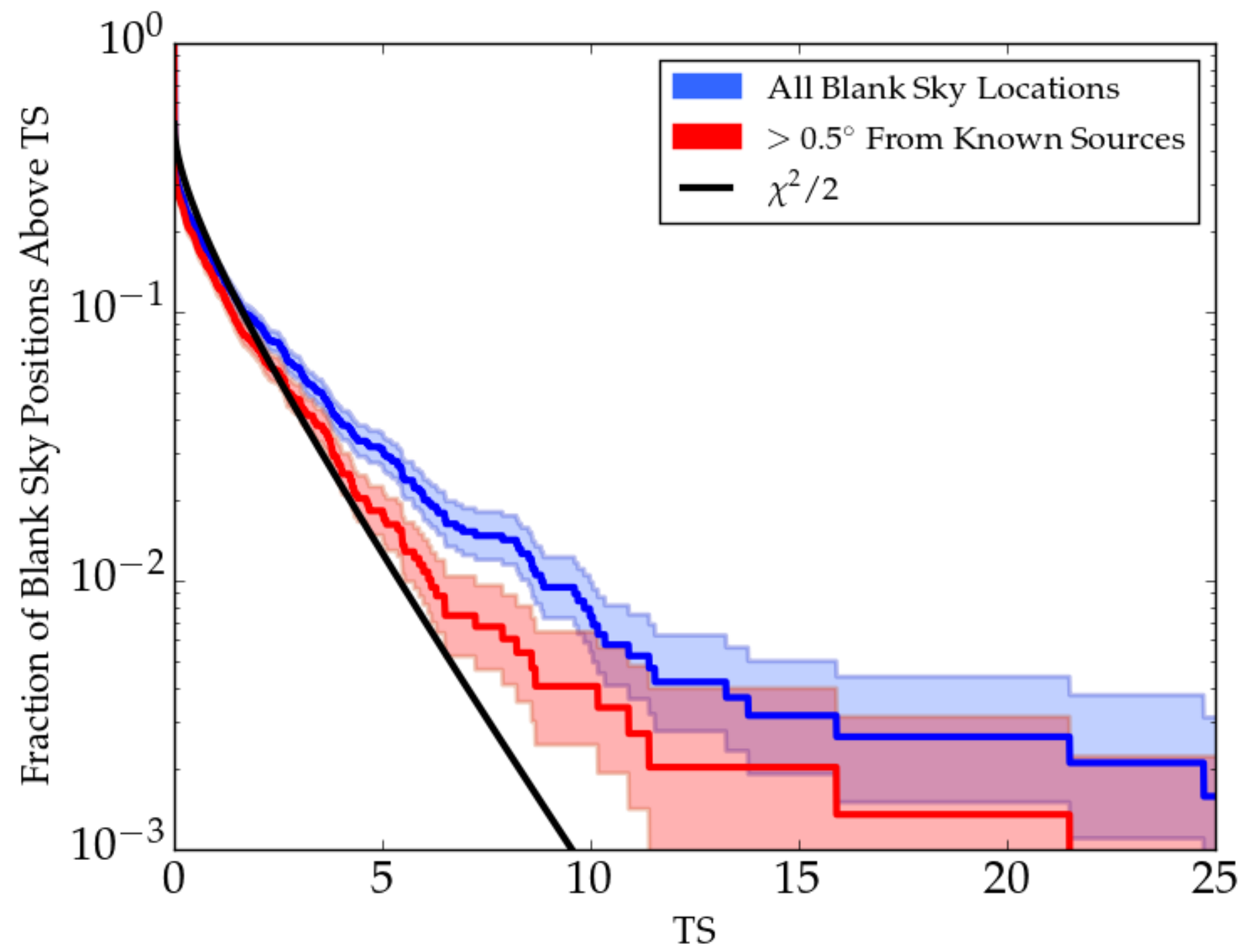,width=0.5\textwidth}   
\caption{\label{fig:blanksky} The fraction of ``blank sky'' locations with a test statistic (TS) larger than a given value, as empirically determined for a collection of 1905 randomly selected sky locations constrained to lie at a galactic latitude $|b|>30^{\circ}$ and at least $1^{\circ}$ ($5^{\circ}$) from point-like (extended) 3FGL sources~\cite{TheFermi-LAT:2015hja}. For the blue curve, no additional requirements are placed on the blank sky locations. For the red curve, the blank sky locations used are additionally required to lie no closer than $0.5^{\circ}$ from any source listed in the BZCAT, CRATES, CGRaBS, or ATNF catalogs (see Sec.~\ref{blanksky}). The shaded region surrounding each curve represents the poisson errors on this determination. In generating this figure, we have adopted a spectral shape corresponding to a 49 GeV dark matter particle annihilating to $b\bar{b}$ (corresponding to the best-fit mass for the Galactic Center gamma-ray excess~\cite{Calore:2014xka}).}
\end{figure}

\begin{figure}
\epsfig{file=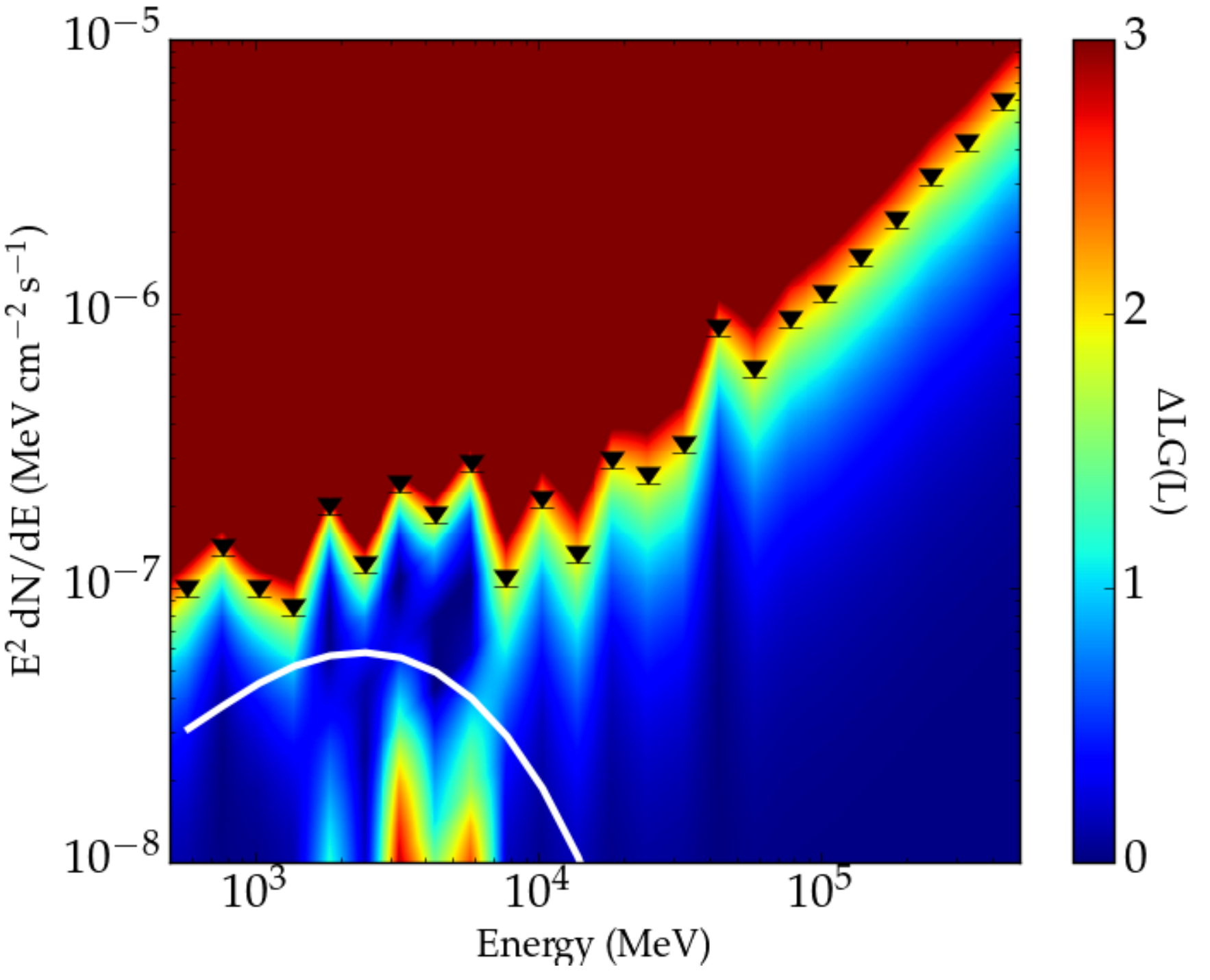,width=0.5\textwidth}   
\caption{\label{fig:ret2} The log-likelihood fit of Reticulum II in 24 energy bins spanning 500
MeV to  500 GeV. The upper limits correspond to 2$\sigma$ confidence in
each energy range. The white line corresponds to the best fit from a
49 GeV dark matter particle annihilating to $b\bar{b}$.}
\end{figure}

Secondly, we apply the ``blank-sky'' null-test employed in previous dwarf spheroidal studies. Specifically, we select 1905 sky locations with $|b|$~$>$~30$^\circ$, which are 1$^\circ$ removed from any 3FGL source and 5$^\circ$ removed from any extended 3FGL source. In this case, we employ the full 6.5~years of data, adopt the default energy range, and test the comparison to a 49~GeV dark matter model annihilating to $b\bar{b}$ (corresponding to the best-fit value of the mass for the spectrum of the Galactic Center excess~\cite{Calore:2014xka}). In Fig.~\ref{fig:blanksky} we show the resulting distribution of our blank-sky test locations. While the existence of systematic errors in the modeling of the gamma-ray background drives this distribution far from that expected from Poisson variations, the result is in good agreement with all previous studies. In this figure, we show results corresponding to the case in which no additional requirements are placed on the blank sky locations (blue), and to when the blank sky locations used are further required to lie no closer than $0.5^{\circ}$ from any source listed in the BZCAT, CRATES, CGRaBS, or ATNF catalogs (red). This will be discussed in more detail in Sec.~\ref{blanksky}.

\begin{table*}[t]
\footnotesize
\begin{tabular}{|c|c|c|c|c|c|c|c|c|}
\cline{4-9}
 \multicolumn{3}{c|}{} &  \multicolumn{2}{c|}{$\log_{10} (J)$} & \multicolumn{2}{c|}{TS (Point-Like)} & \multicolumn{2}{c|}{TS (NFW-Like)} \\
\hline
Dwarf Name   &     Distance (kpc) & Latitude ($^{\circ}$) & Ref.~\cite{Martinez:2013els} & Ref.~\cite{Geringer-Sameth:2014yza} & $m_{\rm DM} =49$ GeV & any $m_{\rm DM}$ & $m_{\rm DM} =49$ GeV & any $m_{\rm DM}$  \\
\hline \hline
Reticulum II  & 32 (30) & -49.7& -- &-- & 17.4 & 18.1 & --&--\\ 
Tucana II   & 58 (69) & -52.4& -- &-- & 1.44 & 1.82 & --&--\\
Indus I   & 69 (100) & -42.1& --& -- & 0.0 & 0.0 &--&--\\
Horologium I   & 87 (79) & -54.7& -- &-- & 0.09& 0.17 &--&--\\
Phoenix II   & 95 (83) & -59.7& --& -- & 0.0 & 0.55 &--&--\\
Eridanus III   & 95 (87) & -59.6& -- &-- & 0.0 & 0.53 &--&--\\
Pictoris I   & 126 (114) & -40.6& --& -- & 0.0 & 0.0 &--&-- \\
Grus 1 & (120) & -58.8 & -- &-- & 0.0 & 0.40 &--&--\\
Eridanus II   & 330 (380) & -51.6& -- &-- & 0.0 & 0.61 &--&--\\
\hline
Triangulum II  & 30 & -23.8& --& --& 0.0& 0.0& --&--\\
\hline
Canis Major   &   7 & -8.0 & --  & -- & 0.0 & 0.0 &--&--\\
Segue 1         &   23  & 50.4 & 19.5 $\pm$ 0.29  & $19.36^{+0.32}_{-0.35}$  & 1.07 & 1.18 &1.10&1.72\\
Sagittarius      &   26  & -14.2 & --  & --  & 2.13 & 4.33 &--&--\\
Ursa Major II         &   32  & 37.4 & 19.3 $\pm$ 0.28 & $19.42^{+0.44}_{-0.42}$ & 0.0 &0.32&0.02&0.88\\
Segue 2         &   35  & -38.1 & --  & $16.21^{+1.06}_{-0.98}$ & 0.49 & 2.09&--&--\\
Willman 1         &   38  & 56.8 & 19.1 $\pm$ 0.31  & -- & 3.94 & 4.47&5.70&5.89\\
 Bootes II   &   42 & 68.9 & --  &-- & 0.0 & 0.25&--&--\\
 Coma Berenices   &  44  & 83.6 & 19.0 $\pm$ 0.25  & $19.02^{+0.37}_{-0.41}$ & 0.0 & 0.0&0.0&0.04\\
Bootes III   &   47 & 75.4 & --  &-- & 0.0 & 0.31 &--&--\\
 Bootes I   &   66 & 69.6 & 18.8 $\pm$ 0.22 & $18.24^{+0.40}_{-0.37}$  & 0.0 & 0.75& 0.09 & 0.50\\
Draco       &  76  & 34.7 & 18.8 $\pm$ 0.16   & $18.84^{+0.12}_{-0.13}$ & 0.04 & 0.05&0.01&0.01\\
Ursa Minor         &   76  & 44.8 & 18.8 $\pm$ 0.19  &  $18.93^{+0.27}_{-0.19}$ & 0.0 & 1.36&0.0&0.99\\
 Sculptor         &   86  & -83.2 & 18.6 $\pm$ 0.18 & $18.54^{+0.06}_{-0.05}$ & 0.01 & 0.02&1.22& 1.28\\
Sextans         &   86  & 42.3 & 18.4 $\pm$ 0.27 & $17.52^{+0.28}_{-0.18}$ & 0.0 & 0.26&0.0&0.24\\
Ursa Major I         &   97  & 54.4 & 18.3 $\pm$ 0.24  & $17.87^{+0.56}_{-0.33}$ & 0.0 & 0.07&0.0&0.24\\
 Carina   &   105 & -22.2 & 18.1 $\pm$ 0.23 & $17.87^{+0.10}_{-0.09}$ & 0.0 & 0.03& 0.0& 0.20\\
Hercules   &  132  & 36.9 & 18.1 $\pm$ 0.25  & $16.86^{+0.74}_{-0.68}$ & 3.09 & 4.05&3.58&4.11\\
Fornax      &  147  & -65.7 & 18.2 $\pm$ 0.21  & $17.83^{+0.12}_{-0.06}$ & 0.36 & 0.94& 1.01& 1.40\\
Leo IV         &   154  & 56.5 & 17.9 $\pm$ 0.28 &  $16.32^{+1.06}_{-1.69}$ & 0.0 & 0.0&0.0&0.0\\
Canes Venatici II   &   160 & 82.7 & 17.9 $\pm$ 0.25 &  $17.65^{+0.45}_{-0.43}$ & 0.27 & 1.56&0.50&1.62\\
Leo V         &   178  & 58.5 & --   &$16.37^{+0.94}_{-0.87}$ & 0.0 & 0.39&--&--\\
Pisces II      &   182  & -47.1 & --   &-- & 0.0 & 0.0&--&--\\
Canes Venatici I   &   218 & 79.8 & 17.7 $\pm$ 0.26 & $17.43^{+0.37}_{-0.28}$ & 0.39 & 0.47&0.35&0.42\\
Leo II         &   233  & 67.2 & 17.6 $\pm$ 0.18 &  $17.97^{+0.20}_{-0.18}$ & 0.0 & 0.0&0.0&0.0\\
Leo I         &   254  & 49.1 & 17.7 $\pm$ 0.18  & $17.84^{+0.20}_{-0.16}$ & 0.0 & 1.67 &0.0& 1.91\\
\hline \hline
\end{tabular}
\caption{The distance, galactic latitude, $J$-factors, and test statistic (TS) of any gamma-ray excess from DES' nine newly discovered dwarf galaxy candidates (top), the new Pan-STARRS dwarf candidate (middle), and the 25 previously known Milky Way dwarf galaxies (bottom). For each of the DES dwarf candidates, we list distances as reported in Ref.~\cite{Bechtol:2015wya} (and as reported in Ref.~\cite{Koposov:2015cua}). For Triangulum II, we list the distance as reported in Ref.~\cite{PanSTARRS}. For the known dwarf galaxies, distances are as given in Refs.~\cite{Martinez:2013els,Geringer-Sameth:2014yza}. Although we list only central values for distances, the error bars on these quantities are typically on the order of $\pm (10-15)\%$. 
The $J$-factors are averaged over a $0.5^{\circ}$ radius around each dwarf, as reported by Refs.~\cite{Martinez:2013els} and \cite{Geringer-Sameth:2014yza}, respectively, and are given in units of GeV$^2$cm$^{-5}$. The TS values listed for each dwarf assume either a spectral shape that corresponds to a 49 GeV dark matter particle annihilating to $b\bar{b}$ (corresponding to the best-fit mass for the Galactic Center gamma-ray excess~\cite{Calore:2014xka}) or allowing the dark matter mass to be a free parameter. For those dwarf galaxies constrained by stellar kinematics, we also show the TS values found when the source is treated as a spatially extended object, corresponding to an NFW halo with a scale radius equal to central value reported in Refs.~\cite{Ackermann:2013yva,Martinez:2013els}. The most significant detection is from the newly discovered and nearby dwarf galaxy candidate Reticulum II.}
\label{tab:TS}
\end{table*}

\section{Results}

In Fig.~\ref{fig:ret2}, we show the delta-log-likelihood ($\Delta LG(L)$) distribution for our analysis of Fermi data from the direction of Reticulum II. As in both Ref.~\cite{Drlica-Wagner:2015xua} and Ref.~\cite{Geringer-Sameth:2015lua}, we find an excess of events in the bins covering approximately $\sim$2-10 GeV. For a spectral shape corresponding to a 49 GeV dark matter particle annihilating to $b\bar{b}$ (the best-fit mass for the Galactic Center excess~\cite{Calore:2014xka}), we find a value of TS=17.4 from Reticulum II, corresponding to a significance of 3.2$\sigma$ (see Fig.~\ref{fig:blanksky}). If we do not impose this choice of the dark matter mass, but rather allow the mass to float as a free parameter, the value of the TS increases only slightly (to 18.1), illustrating the compatibility between this signal and that observed from the Galactic Center.

In Table~\ref{tab:TS}, we list the TS values found in our analysis for each of the previously known Milky Way dwarf spheroidal galaxies, and for the ten newly discovered dwarf galaxy candidates.  Values are given assuming either a spectrum corresponding to the best-fit mass for the Galactic Center excess, or for any dark matter mass. For a Galactic Center-like spectrum, Reticulum II yields the highest significance (TS=17.4), followed by Willman 1 (3.94), Hercules (3.09), Sagittarius (2.13), Tucana II (1.44), and Segue 1 (1.07). Other than that from Reticulum II, no statistically significant excesses are observed.

\begin{figure*}
\epsfig{file=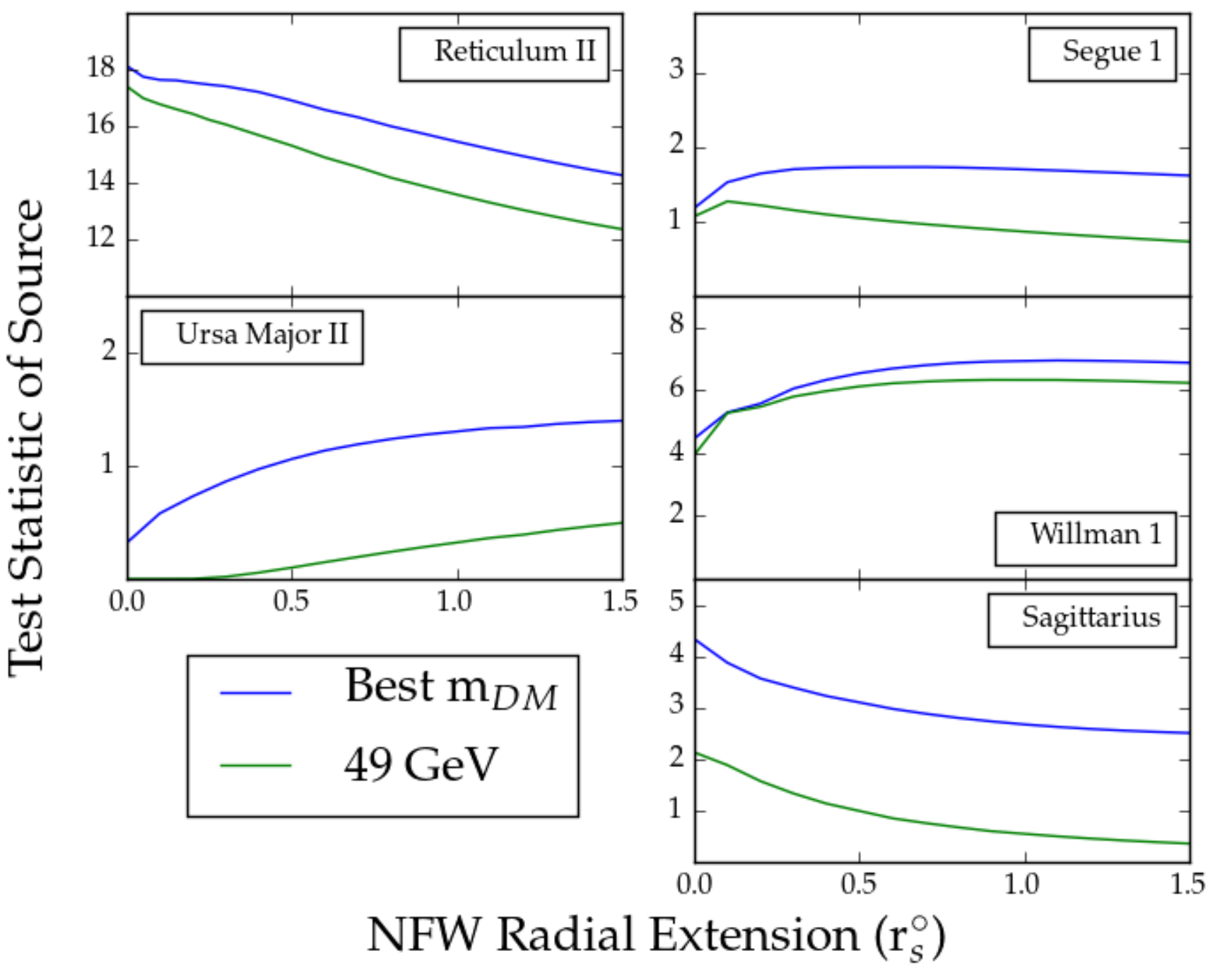,width=0.85\textwidth}   
\caption{\label{fig:ext} The test statistic (TS) as a function of the assumed scale radius of the (NFW) dark matter halo (in degrees, $\tan \theta_{s} \equiv r_s/D$) for five selected dwarf galaxies. The blue (green) curves correspond to a spectral shape for the best-fit dark matter mass (for $m_{\rm DM} =49$ GeV) annihilating to $b\bar{b}$.}
\end{figure*}

Also given in Table~\ref{tab:TS} are the values of the $J$-factors for each dwarf galaxy with sufficient kinematic information (from spectroscopic data) to obtain a determination. This quantity is defined as follows: 
\begin{equation}
J=\int_{\Delta \Omega} \bigg[ \int_{\rm los} \rho^2 dl  \bigg],
\end{equation}
where $\Delta \Omega$ is taken to be a circle of $0.5^{\circ}$ radius around the given dwarf, $\rho$ describes the dark matter density profile of the dwarf, and the second integral is performed over the observed line-of-sight (los). We provide the $J$-factor values as reported by two groups: Martinez {\it et al.}~\cite{Martinez:2013els} and Geringer-Sameth {\it et al.}~\cite{Geringer-Sameth:2014yza}. In general, the dark matter profiles of the classical dwarfs are well constrained by stellar kinematics, resulting in relatively well determined $J$-factors. In contrast, the ultra-faint dwarfs (Segue 1, Ursa Major II, Willman 1) contain far fewer stars, and exhibit much larger $J$-factor error bars. Deeper measurements, capable of detecting more numerous faint stars, will ultimately improve this situation.  Although no spectroscopic information exists for any of the ten new dwarf galaxy candidates, we expect such follow-up measurements to occur in the near future.  

For those dwarfs with profiles constrained by stellar kinematics, we also list in Table~\ref{tab:TS} the values of the TS found when the source is treated as a spatially extended object, rather than as a point-like source. In particular, we adopt an NFW-profile for these systems, with a scale radius equal to the central value reported in Refs.~\cite{Ackermann:2013yva,Martinez:2013els}. In Fig.~\ref{fig:ext}, we plot the TS as a function of the halo's scale radius (in degrees, $\tan \theta_{s} \equiv r_s/D$) for five of the dwarf galaxies under consideration. No strong evidence for (or against) spatial extension is observed. The significances of Willman 1 and Ursa Major II marginally increase if an extended halo is assumed, while the significances of Reticulum II and Sagittarius marginally decrease.

\section{Controlling Backgrounds With Multi-Wavelength Source Catalogs}
\label{blanksky}

In Ref.~\cite{Carlson:2014nra}, it was pointed out that Fermi's sensitivity to dark matter annihilation in dwarf spheroidal galaxies could be increased by taking into account information available in multi-wavelength source catalogs. In particular, a significant fraction of the highest TS points in the ``blank sky'' correspond to the locations of unresolved blazars, radio galaxies, and starforming galaxies.  By making use of only regions of the ``blank sky'' which are not near sources listed in multi-wavelength catalogs, it is possible to reduce the contamination from such sources.  

In Fig.~\ref{fig:blanksky}, the TS distribution of the high-latitude blank-sky is shown without utilizing multi-wavelength information (blue), and after avoiding all locations located within $0.5^{\circ}$ of any source listed in the Roma-BZCAT Multi-Frequency Catalog of Blazars (BZCAT)~\citep{Massaro:2008ye}, the Combined Radio All-Sky Targeted Eight-GHz Survey (CRATES) catalog~\cite{Healey:2007by}, the Candidate Gamma-Ray Blazar Survey (CGRaBS) catalog~\cite{Healey:2007gb}, or the Australia Telescope National Facility (ATNF) pulsar catalog~\cite{Hobbs:2003gk} (red). The application of this cut significantly reduces the fraction of the sky with large TS values.  

To take this multi-wavelength information into account, we apply the following procedure in our analysis. For a given dwarf galaxy (or dwarf galaxy candidate), we check the catalogs described in the previous paragraph for any sources located within $0.5^{\circ}$. If any are found, we re-run our analysis, including in the background model a source at that location. We then take the new TS of the dwarf, and see how many locations on the blank sky yield a higher TS, when a background source is included at the location of the nearby catalog source. We then use the fraction of high-TS blank-sky locations to calculate the $p$-value and significance of any dwarf galaxy excess. 

This procedure is most important in the case of Reticulum II, which is located 0.44$^{\circ}$ from the source CRATES J033553-543025.\footnote{The presence of this source was pointed out to us by Eric Carlson.} Given the large number of sources contained in these catalogs, this is not a particularly surprising (we estimate a probability of $\sim$20\% that at least one source would reside within $0.5^{\circ}$). The analysis with the extra background source at the CRATES location resulted in a TS of 13.4 from Reticulum II (a modest reduction from 17.4).\footnote{We note that the spectral shape absorbed by the CRATES source is very hard, and quite unlike that of the gamma-ray emission from typical radio sources. We consider it unlikely that this source contributes significantly to the gamma-ray flux observed from the direction of Reticulum II.} After re-running our analysis on all ``blank-sky'' locations with TS>13.4, including in the background model sources at the locations of the multi-wavelength catalog sources, we find that only 3 out of 1905 of the blank-sky locations yielded a more significant excess. This corresponds to a $p-$value of 0.001575 and a detection significance of 3.2$\sigma$ (compared to 3.0$\sigma$, which is found if no multi-wavelength information is utilized). If we include the poisson error bars around the number of 3 blank-sky locations, the corresponding significance covers the range of 3.0 to 3.4$\sigma$.

\begin{figure}
\epsfig{file=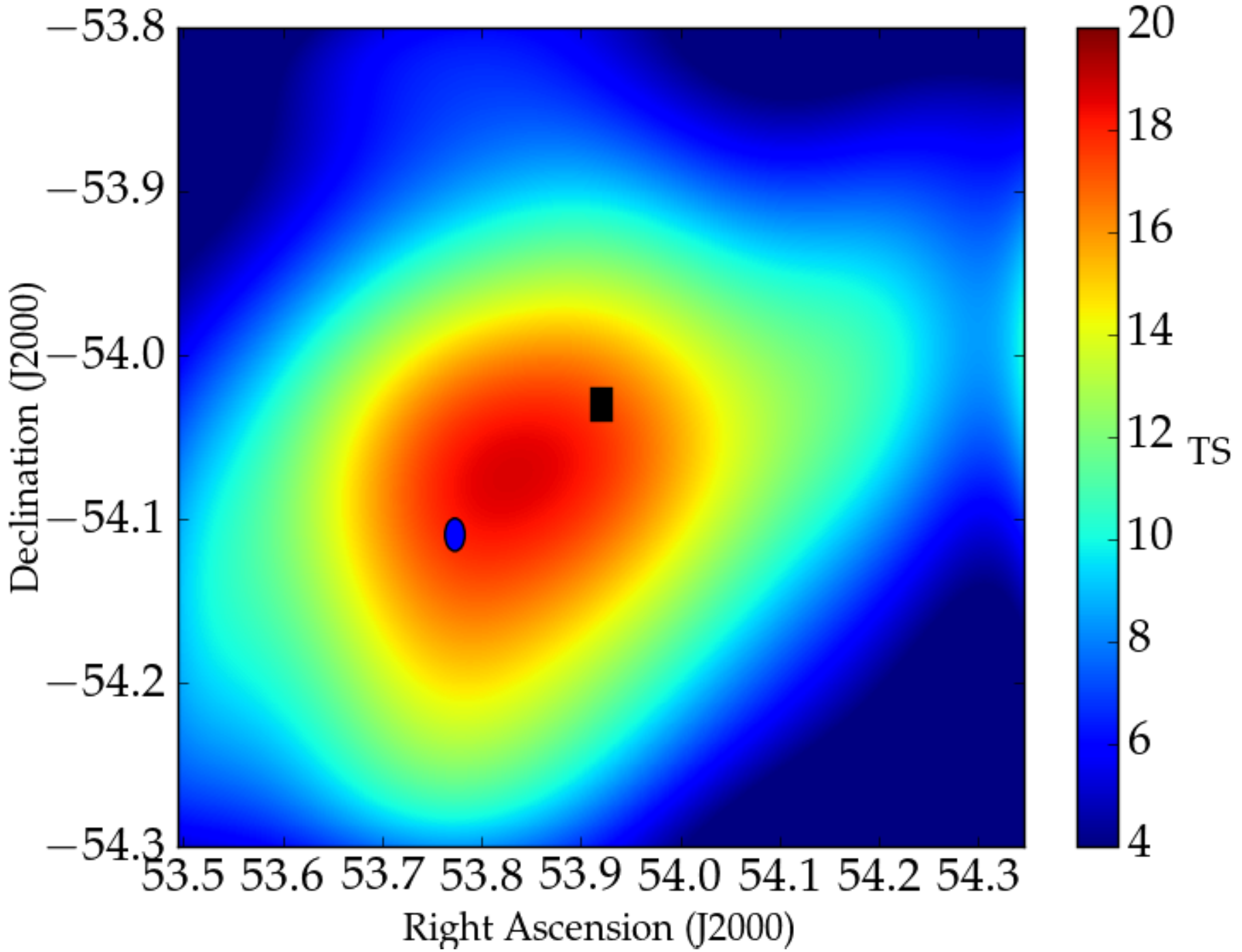,width=0.5\textwidth}   
\caption{\label{retsky} The test statistic (TS) over the region of the sky surrounding Reticulum II. Although the excess gamma-ray emission is localized to a region of approximately $\sim0.1^{\circ}$ radius, it is not possible to spatially discriminate Reticulum II (black square) from the location of the radio source PMN J0335-5406 (blue oval).}
\end{figure}

As we were finalizing this paper, it was pointed out to us that the faint radio source PMN J0335-5406 in the Parkes-MIT-NRAO catalog is located $0.105^{\circ}$ from the location of Reticulum II.\footnote{We thank Alex Drlica-Wagner for bringing this source to our attention.}  The PMN catalog contains a much larger number of sources (50,814) than the catalogs employed in our analysis, and it is not particularly surprising that a such source resides near Reticulum II (or near any other location in the high-latitude sky). Specifically, at the declination of Reticulum II, the PMN survey contains approximately 17 sources per square degree, corresponding to a 17\% chance that a catalog source exists within a 0.1$^\circ$ ROI around Reticulum II. In Fig.~\ref{retsky}, we plot the value of the TS found by our analysis for regions of the sky in the vicinity of Reticulum II. Although we can localize the excess gamma-ray emission to a region of approximately $\sim$$0.1^{\circ}$ radius, it is not possible to spatially discriminate Reticulum II (black square) from the location of the PMN source (blue oval). The PMN source exhibits continuous radio emission that is consistent with a blazar origin. When re-running our analysis for a power-law gamma-ray spectrum (as is observed from blazars) we find a best-fit spectral index of 2.04, yielding TS=13.1. In comparing this to the TS value found earlier in this study, we conclude that a dark matter-like spectrum is preferred to a blazar-like power-law at the level of approximately $\sim 2\sigma$. Spectroscopic observations of this source would be valuable, as they could aid in determining its nature and inform us as to its likely gamma-ray luminosity and spectrum. 

\section{A Self-Consistent Interpretation}

In this section, we consider the excess observed from Reticulum II, along with the lack of significant detections from other Milky Way dwarf galaxies, and ask whether these results are mutually consistent. Focusing on the case in which annihilating dark matter is responsible for the Galactic Center excess ($m_{\rm DM} = 49$ GeV, for the case of annihilatiions to $b\bar{b}$), the lack of significant excess emission from the known dwarf galaxies constrains $\sigma v \lsim 1.3 \times 10^{-26}$ cm$^3$/s~\cite{Ackermann:2015zua}. And while this constraint is compatible with dark matter interpretations of the Galactic Center excess, the normalization of the Galactic Center signal implies that the cross section is unlikely to be smaller than this value by more than a factor of a few. More specifically, if we allow the overall normalization and the scale radius of the Milky Way's dark matter halo profile to vary within the range allowed by dynamical constraints ($\rho_{\rm local} = 0.24-0.46$ GeV/cm$^3$, $r_s=8-35$ kpc~\cite{2011JCAP...11..029I,2010JCAP...08..004C}), we find consistency with an annihilation cross section as small as $\sim 3\times 10^{-27}$ cm$^3$/s. From this perspective, the prospects for the future detection of a gamma-ray signal from one or more dwarf galaxies appears encouraging. 

For an annihilation cross section at the upper limit of Ref.~\cite{Ackermann:2015zua} ($\sigma v \simeq 1.3 \times 10^{-26}$ cm$^3$/s), the normalization of the signal from Reticulium II requires approximately $\log J \simeq 19.6$--20.1. Noting the empirical (and approximate) relationship between the distances and $J$-factors of ultra-faint dwarfs, Ref.~\cite{Drlica-Wagner:2015xua} points out that Reticulium II might be expected to have a somewhat smaller value, $\log J \sim 19.3$, although even a value as high as of 20.1 would not be a particuarly significant outlier. The necessity of a large $J$-factor for Reticulum II (if its gamma-ray excess is from annihilating dark matter) can also be seen from the results of our analysis, as shown in Table~\ref{tab:TS}. Roughly speaking, the predicted value for the TS of a given dwarf is proportional to its gamma-ray flux, and thus to its $J$-factor. The modest TS values observed from Segue 1 and Ursa Major II suggest significantly lower $J$-factors for these systems than for Reticulum II. Given this situation, we eagerly await the spectroscopic follow-up of Reticulium II.  If the gamma-ray excess from this source in fact originates from annihilating dark matter, we should anticipate a large value for its $J$-factor, likely in excess of $\sim$$10^{19.6}$ GeV$^2$ cm$^{-5}$.



\section{Are Statistical Fluctuations Sufficient To Explain The Differences Between The Results Found Using Pass 7 and Pass 8 Data?}

At face value, our determination of TS=18.1 from the direction of Reticulum II appears to be in conflict with the more modest value of 6.7 quoted by the Fermi Collaboration~\cite{Drlica-Wagner:2015xua}. The most significant difference between these two analyses is in the data sets that are being considered: our analysis makes use of the publicly available Pass 7 data, whereas the Fermi Collaboration paper utilizes the more recent Pass 8 data set. Intriguingly, a similar discrepancy can be seen in a comparison of the Fermi's Collaboration's Pass 7 and Pass 8 studies of known dwarf galaxies. In particular, whereas the Fermi Collaboration's Pass 7 analysis revealed an excess from three ultra-faint dwarfs (Segue 1, Ursa Major II, and Willman 1), at a level of TS $\approx 10$~\citep{Ackermann:2013yva}, no excess was observed according to their more recent Pass 8 paper~\cite{Ackermann:2015zua}. This would-be conflict between the Fermi Pass 7 and Pass 8 dwarf papers is somewhat surprising in light of the fact that these two analyses make use of data taken over significantly overlapping time periods (over the first four years of Fermi's mission; the more recent analysis adds two more years of data to this set). The Fermi Collaboration points out, however, that after taking into account the new event selection associated with the transition from Pass 7 to Pass 8, the overlap between these two data sets is not particularly large; only $\sim$30\% of the 1-10 GeV photons used in the most recent analysis were also employed in their earlier study (put another way, approximately 65\% of those events in Fermi's Pass 7 dwarf analysis were also included in the recent Pass 8 study)~\cite{Ackermann:2015zua}.  Assuming that the previous TS $\approx$ 10 excess was the result of a statistical fluctuation, we estimate that there was an approximately $\sim$10\% chance that the more recent data set would yield TS $<1$ (after taking into account differences in effective area and exposure time). So in this respect, we concur with the conclusion of the Fermi Collaboration. 

In order to compare our results directly to those of Ref.~\citep{Ackermann:2013yva}, we reanalyzed the three ultra-faint dwarf spheroidal galaxies (Segue 1, Willman 1, and Ursa Major II) using the P7V6 Reprocessed data, employing in this test only four years\footnote{MET range: 239557417 -- 365817602} of data and utilizing only those events which pass the P7V6REP ``Clean'' selection cut (and the  {\tt gll\_iem\_v05.fit} and {\tt iso\_clean\_v05.txt} models for the diffuse and isotropic emission, respectively), and examine the data a in 14$^\circ$~$\times$~14$^\circ$ region around each ultra-faint dwarf in 0.1$^\circ$ spatial bins. Assuming a spectrum corresponding to a 25 GeV dark matter particle annihilating to $b\bar{b}$, we find that these three dwarfs acquire TS values of 1.18, 2.43, and 2.43, respectively.  If we model these sources as extended objects (NFW-like) rather than as point sources, we find TS values of 0.89, 3.05, and 2.73. In either case, our TS values are similar to but somewhat lower than the TS $\approx$ 10 reported in~Ref.~\citep{Ackermann:2013yva}. Comparing this to the results shown in Table~\ref{tab:TS}, however, we see little indication that the addition of 2.5 years of data has reduced the overall TS from these three ultra-faint dwarfs (Ursa Major II's TS fell, while that of Willman 1 increased by a similar amount).  In light of this somewhat confusing situation, we eagerly await the public availability of Pass 8 data.

\section{Summary and Conclusions}

In this article, we have revisited the gamma-ray emission from known Milky Way dwarf galaxies, and from the dwarf galaxy candidates recently discovered in the data from DES and Pan-STARRS.  Of particular interest are the new dwarf candidates Reticulum II and Triangulum II, which are each located at a distance of only $\sim$30 kpc from Earth, making them promising targets for dark matter searches. Our analysis of Fermi data from the direction of Reticulum II identifies an excess of gamma-rays with a local statistical significance of 3.2$\sigma$. This is slightly higher than, but not dissimilar to, that reported in the previous studies of other groups~\cite{Drlica-Wagner:2015xua,Geringer-Sameth:2015lua}. We also confirm that Reticulum II's gamma-ray excess is most prominent at energies between $\sim$2-10 GeV, in good agreement with the spectral shape of the excess previously reported from the region surrounding the Galactic Center. We do not observe any significant $\gamma$-ray emission from the direction of Triangulum II. 

Looking forward, spectroscopic follow-up of Reticulum II and the other new dwarf galaxy candidates will be important for interpreting this data.  In order for this excess to be compatible with the lack of significant gamma-ray detections from other dwarf galaxies (most importantly, Segue 1 and Ursa Major II), Reticulum II must contain a high density of dark matter, corresponding to $J \gsim 10^{19.6}$ GeV$^2$cm$^{-5}$. A measurement of Reticulum II's $J$-factor that is much smaller than this value would place serious doubt as to any dark matter interpretation of its excess. Additional data from Fermi will also have much to bear on this question. With 50\% more data, such as could be acquired over the next few years, we estimate that the detection of Reticulum II's gamma-ray emission could exceed TS=25, corresponding to approximately 4$\sigma$ significance, and comparable to the threshold for membership in Fermi's point source catalogs.

\bigskip

{\bf Acknowledgements.}
We would like to thank Andrea Albert and Alex Drlica-Wagner for helpful comments and discussions, as well as Eric Carlson for pointing out the proximity of the nearby CRATES source. DH is supported by the US Department of Energy under contract DE-FG02-13ER41958. Fermilab is operated by Fermi Research Alliance, LLC, under Contract No. DE- AC02-07CH11359 with the US Department of Energy. TL is supported by the National Aeronautics and Space Administration through Einstein Postdoctoral Fellowship Award No. PF3-140110.

\bibliography{fermidwarfs2015}

\end{document}